\documentclass[11pt,fleqn,a4paper]{article}

\usepackage{amssymb}
\usepackage{graphicx}
\usepackage{hyperref}
\usepackage[authoryear]{natbib}
\usepackage{natbib}
\usepackage{amsmath}
\usepackage[english]{babel}
\usepackage{amsfonts}
\usepackage{rotating}
\usepackage{lscape}
\usepackage[official]{eurosym}
\usepackage{longtable}
\usepackage[a4paper,left=2.5cm,right=2.5cm,top=3cm,bottom=3cm]{geometry}
\usepackage{xspace}
\usepackage{array}
\usepackage{tcolorbox}
\usepackage{latexsym}
\usepackage{float}
\usepackage{bm}

\usepackage{listings}
\usepackage{xcolor}
\usepackage{booktabs}
\usepackage{fancyref}
\usepackage{tkz-graph}
\usetikzlibrary{shapes.geometric}
\usepackage{neuralnetwork}
\usetikzlibrary{arrows.meta}

\usepackage{setspace}%%
\usepackage{xspace}%% für besseres Handling der spaces
\usepackage{authblk}

\usepackage{soul}
\usepackage{chngcntr}
\begin{document}

\begin{center}
	
	{\LARGE An Introduction to Causal Discovery}
	
	{\large \vspace{0.4cm}}
	
	{\Large Martin Huber}\medskip
	
	{ {University of Fribourg, Department of Economics} \smallskip }
\end{center}

{\bf Abstract:} {\small In social sciences and economics, causal inference traditionally focuses on assessing the impact of predefined treatments (or interventions) on predefined outcomes, such as the effect of education programs on earnings. Causal discovery, in contrast, aims to uncover causal relationships among multiple variables in a data-driven manner, by investigating statistical associations rather than relying on predefined causal structures. This approach, more common in computer science, seeks to understand causality in an entire system of variables, which can be visualized by causal graphs. This survey provides an introduction to key concepts, algorithms, and applications of causal discovery from the perspectives of economics and social sciences. It covers fundamental concepts like d-separation, causal faithfulness, and Markov equivalence, sketches various algorithms for causal discovery, and discusses the back-door and front-door criteria for identifying causal effects. The survey concludes with more specific examples of causal discovery, e.g.\ for learning all variables that directly affect an outcome of interest and/or testing identification of causal effects in observational data.}
{\large \vspace{0.4cm}}\\
{\bf Keywords:} {\small Causal discovery, causal inference, causal graphs, d-separation, back-door criterion}
{\large \vspace{0.4cm}}\\
{\footnotesize Address: Martin Huber, Boulevard de P\'erolles 90, 1700 Fribourg, Switzerland. E-mail: martin.huber@unifr.ch.}

\thispagestyle{empty}
\newpage
\setcounter{page}{1}

\section{Introduction}

In socials sciences and economics, causal inference, also known as treatment, program, or impact evaluation, predominantly focuses on evaluating the causal effect of a specific treatment variable, such as a education program, health treatment, or marketing intervention, on an outcome of interest, such as earnings, health, or sales. Comprehensive surveys on causal inference are for instance provided in \cite*{Im04}, \cite*{ImWo08}, and \cite*{AbadieCattaneo2018}, as well as in the textbooks \cite*{angrist2009mostly}, \cite*{frsp2019}, \cite*{Cunningham2021}, \cite*{HuntingtonKlein2022}, and \cite*{huber2023causal}. More recently, causal inference has been combined with machine learning, a subfield of artificial intelligence, which permits considering observed covariates, such as socio-economic characteristics, in a data-adaptive manner, to control for confounding factors that jointly affect the treatment and the outcome (and thus, bias causal estimates) and/or to assess whether effects are heterogeneous across groups with different covariate values. The survey papers by \cite*{lieliuse2022} and \cite*{lechner2023causal} as well as the textbooks by \cite*{huber2023causal} and \cite*{Chernozhukovetal2022} provide an introduction to such causal machine learning methods. 

In contrast to causal inference or causal machine learning for assessing the impact of a predefined treatment variable on a predefined outcome, causal discovery, which is more prominent in computer science than in economics and social sciences, aims to learn the causal relationships among several or even many variables in a data-driven manner. In other words, causal discovery seeks to understand or unveil the causal associations within an entire system of variables, which can be depicted by a causal graph. The task of determining which variables influence others, based solely on statistical associations rather than presupposed causal structures (like assuming that a treatment may affect an outcome but not vice versa), poses a significant challenge in empirical settings. A growing number of studies demonstrate the circumstances and assumptions under which this endeavor is at least theoretically feasible, see for instance the literature reviews by \cite*{KalischB2014}, \cite*{peters2017elements}, and \cite*{Glymouretal2019}. 

This survey provides an introduction to concepts, algorithms, and examples of causal discovery through the lens of economics and social sciences. Section \ref{dsep} discusses fundamental concepts like d-separation, causal faithfulness, and Markov equivalence. Section \ref{causaldisalg} sketches various algorithms for causal discovery. Section \ref{backfront} discusses the back-door and front-door criteria for the identification of causal effects of predefined treatments. Section \ref{examples} concludes with examples of causal discovery for learning all variables that directly affect an outcome of interest and/or testing identification of causal effects in observational data. Section \ref{conclusion} concludes.

\section{d-Separation, Causal Faithfulness, and Markov equivalence}\label{dsep}

\begin{figure}[!htp]
	\centering \caption{\label{DAGs}  Selection on observables (left) and Y-learning (right)}\bigskip
	\begin{tikzpicture}[scale=1.3]
		\SetGraphUnit{2}
	    \Vertex{D}  \EA(D){Y}  \SO(D){X}
        \Edges(D,Y) \Edges(X,D) \Edges(X,Y)
	\end{tikzpicture}$\hspace{6em}$
	\begin{tikzpicture}[scale=1.3]
	\SetGraphUnit{2}
	\Vertex{D}  \EA(D){Y}  \SOWE(D){Z} \NOWE(D){X}
	\Edges(D,Y) \Edges(X,D) \Edges(Z,D)
\end{tikzpicture}
\end{figure}
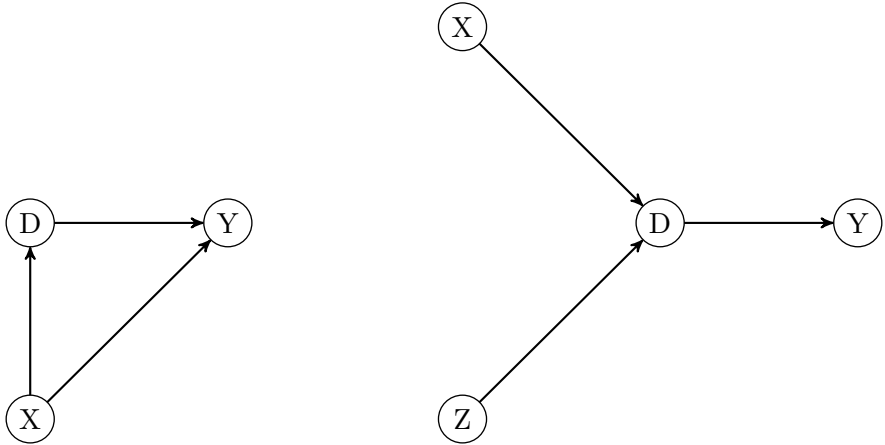

This section introduces essential concepts in causal discovery: d-separation, causal faithfulness, and Markov equivalence. D-separation, short for ``dependency separation'', establishes a formal framework to analyze and comprehend the relationships between variables within causal graphs. Causal graphs describe a causal system of variables and consist of nodes, which represent individual variables or sets of variables, and arrows, which indicate causal effects between these variables. For instance, an arrow originating from a node (or variable) $D$ and pointing to node $Y$ signifies that variable $D$ has a causal impact on variable $Y$, as illustrated in the left graph of figure \ref{DAGs}. As an example, consider education ($D$) having a causal effect on earnings ($Y$). Furthermore, directed acyclic graphs (DAGs) rule out cyclic or simultaneous relations, like arrows going both from $D$ to $Y$ and $Y$ to $D$. It is therefore ruled out that education ($D$) simultaneously affects and is affected by earnings ($Y$) at the point in time when both variables are measured. D-separation enables the identification of conditional independence relationships among variables based on the structure of a DAG. 

The d-separation criterion of  \cite*{pearl1988probabilistic} is founded on the concept of ``blocking'' or controlling for causal paths in a graph in a manner that establishes statistical (conditional) independence between two variables. This implies that after blocking specific paths connecting the two variables, they are no longer associated with each other. More concisely, a path between two (sets of) variables $D$ and $Y$ is blocked when conditioning on a (set of) control variable(s) $C$,  
\begin{enumerate}
	\item if the path between $D$ and $Y$ is either a causal chain, implying that $D\rightarrow X \rightarrow Y$ or $D\leftarrow X \leftarrow Y$, or a confounding association, implying that $D\leftarrow X \rightarrow Y$, and variable (set) $X$ is among control variables $C$ (i.e.\ controlled for),
	\item if the path between $D$ and $Y$ contains a collider, implying that $D\rightarrow  S  \leftarrow Y$, and variable (set) $S$ or any variable (set) causally affected by $S$ is not among control variables $C$ (i.e.\ not controlled for).
\end{enumerate}

With this definition of blocking in mind, the d-separation criterion states that variables $D$ and $Y$ are d-separated when conditioning on control variable(s) $C$ if and only if $C$ blocks all paths between $D$ and $Y$.  In other words, $D$ and $Y$ are d-separated (i) if we control for all (mediating) variables through which $D$ affects $Y$ or vice versa, as well as all (confounding) variables that jointly affect $D$ and $Y$, and (ii) if we do not control for any (collider) variables that are jointly affected by $D$ and $Y$, or any variables influenced by colliders. Since d-separation is a sufficient condition for the (conditional) independence of two variables, it is very useful for causal reasoning in complex causal models. For this reason, d-separation serves as a theoretical basis for causal discovery algorithms designed to learn causal models from data. Reconsidering the left graph of figure \ref{DAGs}, we observe that $D$ and $Y$ are not d-separated when we control for variable $X$. Despite the fact that the confounding relationship $D\leftarrow X \rightarrow Y$ is blocked when conditioning on $X$, the causal effect $D\rightarrow Y$ remains unblocked such that $D$ and $Y$ are dependent conditional on $X$. As an example, consider the case that education ($D$) and earnings ($Y$) are confounded by background characteristics like socio-economic status or innate ability ($X$). After controlling for such characteristics $X$ to avoid confounding or omitted variable bias, the conditional dependence of $Y$ and $D$ reflects the causal effect of education on earnings under the causal model in the left graph of figure \ref{DAGs}. 

\tikzstyle{EdgeStyle}   = [->,>=stealth']

To illustrate the usefulness of d-separation for causal discovery, consider the causal model depicted in the right graph of figure \ref{DAGs}, which pertains to Y-learning, as e.g.\ discussed in \cite*{mani2012theoretical}. Variables $X$ and $Z$ are independent of each other when not controlling for $D$, while they are dependent conditional on $D$. According to the d-separation criterion, this necessarily implies that $D$ is a collider ($X\rightarrow  D  \leftarrow Z$). Furthermore, both $X$ and $Z$ are associated with $Y$, as they both affect $Y$ via $D$. However, when controlling for $D$, $Y$ is conditionally independent of both $X$ and $Z$. According to the d-separation criterion, this necessarily implies that $D$ functions as a mediator, transmitting the causal effects of $X$ and $Z$ to $Y$ ($X\rightarrow  D  \rightarrow Y$, $Z\rightarrow  D  \rightarrow Y$). Put differently, $X$ and $Z$ are so-called instruments for $D$, as they both affect $D$ but do not directly affect $Y$ (other than through $D$), thus, satisfying an exclusion restriction. 

As an example, consider the case that  $D$ is participation in a training, $Y$ reflects earnings after training participation, and $X$ and $Z$ are randomly assigned e-mail- and text message-based invitations to participate in the training. In the absence of confounding and if the invitations affect the earnings outcome only through training participation, then the invitations are independent of earnings conditional on training participation. Furthermore, both types of invitations are dependent of each other conditional on training if both types affect training participation. Our Y-learning example demonstrates that in specific causal models, d-separation may enable the comprehensive learning of the entire causal model, but this is not the case in general and unlikely to apply to  most causal problems. 

While d-separation is sufficient for the conditional independence of two variables, it is important to note that in special cases, two variables might be independent even if d-separation fails. To illustrate this, let us revisit the left graph in figure \ref{DAGs}. If the confounding path of $D$ and $Y$ due to $X$ ($D\leftarrow X \rightarrow Y$)  perfectly offsets the causal effect of $D$ on $Y$ ($D\rightarrow Y$), then $D$ and $Y$ are statistically independent even though d-separation fails (due to the presence of both confounding and the causal effect of $D$ on $Y$). As an example, consider the case that socio-economic status ($X$) affects both participation in a training program ($D$) and earnings ($Y$) in a way that offsets the effect of the training on earnings, such that training and earnings are statistically independent despite the existence of an earnings effect of training. 

Causal discovery methods typically rely on the absence of different paths between variables that exactly cancel each other out. For this reason, we may need to explicitly rule out offsetting paths through the causal faithfulness (or stability) assumption, see for instance the discussions in \cite*{Pearl00} and \cite*{spirtes2000causation}. Causal faithfulness imposes that only variables that are d-separated in a DAG are conditionally independent, while any variables that are not d-separated are dependent. Consequently, the assumption mandates that d-separation is not only a sufficient, but also a necessary condition for conditional independence. In other words, two variables are statistically independent if and only if d-separation is met.

However, even with the application of causal faithfulness, d-separation may not (and in most observational data will not) lead to a unique determination of the causal model that underlies the observed associations of variables in the data. This implies that the conditional dependence and independence relationships detected through the d-separation criterion might be compatible with multiple causal models, resulting in ambiguity regarding the true causal structure. This issue motivates the concept of Markov equivalence classes for characterizing causal models that entail identical conditional dependencies and independencies. Distinct causal models belong to the same Markov equivalence class if they exhibit the exact same patterns of conditional dependence and independence in the data, such that the causal models are indistinguishable based on observational data alone. 

To provide an example, let us consider two distinct causal models. In the first model, $D$ (e.g.\ education) affects $Y$ (e.g.\ earnings) exclusively through a mediator $M$ (e.g.\ human capital acquisition): $D\rightarrow M \rightarrow Y$. In the second model,  $Y$ affects $D$ exclusively through $M$: $D\leftarrow M \leftarrow Y$. In the absence of any further variables, the two models  generate precisely the same patterns of conditional dependence and independence: $D$ and $Y$ are conditionally independent given $M$, while $D$ and $M$ are dependent given $Y$ and $Y$ and $M$ are dependent given $D$, and any of $D,M,Y$ are mutually dependent in the absence of conditioning on any variable. As a result, both models belong to the same Markov equivalence class, because we are unable to distinguish between these models based solely on the statistical associations in the data. Causal discovery becomes even more involved when allowing for unobserved variables in a causal model, e.g.\ an unobserved confounder $U$ jointly affecting $D$ and $Y$ ($D\leftarrow U \rightarrow Y$), such as unobserved personality traits ($U$) affecting both education ($D$) and earnings ($Y$). Since the d-separation criterion cannot be applied to unobserved variables, their presence tends to exacerbate the uncertainty regarding the true causal model. Consequently, this can increase the number of causal models that align with the same Markov equivalence class. 

In scenarios where the correct causal model cannot be learned solely from observational data, the challenge of model ambiguity can potentially be addressed through ``external'' sources of information concerning the causal structure. Such sources may include domain expertise, theoretical insights, past empirical findings, or knowledge about the temporal sequence of events. These external insights can provide valuable context that helps narrow down potential causal models. For example, if variable $D$ (e.g., training participation) is measured at an earlier time point than variable $Y$ (e.g., earnings after training participation), it becomes evident that a causal path from $Y$ to $D$ (such as $D \leftarrow Y$) can be ruled out. This is because future events cannot exert influence on past events, a principle discussed in works like \cite*{reichenbach1991direction} concerning the connection between causality and time. Moreover, in order to investigate whether an observed association between training $D$ and earnings $Y$ is solely due to confounding ($D\leftarrow U \rightarrow Y$), or if it also involves a nonzero treatment effect ($D \rightarrow Y$), an experiment could be conducted in which the treatment is randomly assigned, as random assignment avoids confounding. Consequently, external sources of information might be helpful for clarifying some of the ambiguous causal associations in a DAG in order to reduce the number of plausible causal models within in a Markov equivalence class. 

\section{A Sketch of Algorithms for Causal Discovery}\label{causaldisalg}

As discussed in the previous section,  the objective of causal discovery is to recognize Markov equivalence classes encompassing all causal models that are indistinguishable (or statistically equivalent) in observed data because they exhibit the same d-separation patterns (i.e.\ patterns of conditional independence). For the practical application of the d-separation criterion in causal discovery, \cite*{verma1990equivalence} introduced the IC (or I-equivalence Class) algorithm for observational data. The IC algorithm operates under the assumptions that causal faithfulness holds and that there are no  unobserved confounders that jointly affect any pair of observed variables for which the causal associations are to be estimated. The IC algorithm consists of the following steps:
\begin{enumerate}
 	\item For all pairs of variables $A$ and $B$ in the data, search for a set of variables $X$ such that  $A$ and $B$ are conditionally independent when controlling for $X$. Link $A$ and $B$ by an edge, which represents an undirected association in a causal graph, if and only if no $X$ exists that satisfies conditional independence.
	\item For all pairs of variables $A$ and $B$ that do not share an edge with each other, but  share both an edge with a variable $S$, verify whether $S$ is in set $X$ of step 1. If this is not the case, $S$ is a collider such that the causal association is  $A\rightarrow S\leftarrow B$.
	\item In the resulting graph with partially determined causal associations (directed causal arrows) and partially undetermined associations (edges), orient the direction of as many edges as possible subject to two conditions: (i) Any alternative orientation would yield a new collider structure. (ii) Any alternative orientation would yield a directed cycle (i.e.\ a circular causal relation between variables). 
\end{enumerate}

Bluntly speaking, step 1 of the algorithm finds those pairs of variables that are dependent conditional on any feasible set of control variables. Variables in such a pair are then connected by an undirected edge since the specific causal path remains unknown at this stage. Step 2 pinpoints collider paths which unveil the causal directions (arrows) between variables. Step 3 finds further causal associations that adhere to the constraints of not creating circular causal relations (which are prohibited in DAGs) or unwarranted further collider paths (as correct collider paths were already detected in step 2). As highlighted by \cite*{Pearl00}, step 3 can be systematized in several ways, e.g.\ by applying the rules provided in \cite*{verma1992algorithm} for orienting edges into causal arrows. These rules are sufficient to identify the maximum potential number of causal arrows in a causal graph based on observed data. 

\cite*{spirtes1991algorithm} introduced a refined method known as the PC algorithm. It reduces computation time by limiting step 1 of searching for sets $X$ that entail the conditional independence of $A$ and $B$ to variables that share edges with (i.e., are adjacent to) either $A$ or $B$. \cite*{Glymouretal2019} offer a comprehensive discussion and illustration of the PC algorithm based on the causal scenario of Y-learning considered in section \ref{dsep}. Numerous further enhancements have been proposed in causal discovery algorithms, e.g.\ for integrating external information regarding causal relationships between specific variables. One important contribution is the Fast Causal Inference (FCI) algorithm introduced by \cite*{spirtes2000causation}, which allows for and in some causal models even detects the presence of unobserved variables. \cite*{Glymouretal2019} discuss an example for the application of the FCI algorithm in the presence of unobserved confounders. 

An important question in causal discovery is how to perform statistical inference, e.g.\ to obtain p-values or confidence intervals for the estimated causal effects between variables. To test the various conditional independence assumptions between pairs of variables in the data, algorithms typically adopt either pairwise test statistics like t-tests or global goodness-of-fit statistics that are computed collectively for all variables (rather than pairs of variables), such as the Bayes Information Criterion (BIC). In this context, it is important to acknowledge that verifying conditional independencies among multiple variables amounts to jointly testing several hypotheses and therefore introduces multiple hypothesis testing issues. Consequently, the critical values of pairwise tests must be adjusted to account for the number of tested causal associations (which grow exponentially in the number of variables in the causal model), e.g.\ by a Bonferroni-type adjustment, see e.g.\ \cite*{holm1979simple}. Otherwise, there is an increased risk that testing might (by random chance) erroneously indicate the presence of specific causal associations that are actually absent, implying an increased type I error rate in testing. However, on the flip side, such corrections for multiple hypothesis testing may decrease the power of the tests, implying a reduced chance (or probability) to detect non-zero causal effects, implying an increased type II error rate in testing. This issue becomes particularly pronounced when there are many variables in the causal model, which implies more rigorous corrections. 

The algorithms discussed so far explore conditional independencies and dependencies to estimate the set of causal models that fit within the Markov equivalent class of the true (albeit unknown) causal model. However, there are alternative avenues to causal discovery that rely on specific assumptions about the functional nature of causal relationships between variables. Under these assumptions, it becomes possible to uniquely determine the true causal model without relying on d-separation or causal faithfulness. As discussed in \cite*{hoyer2008nonlinear}, \cite*{zhang2009causality}, and \cite*{peters2014causal}, causal relationships can be learned from the data under the following conditions: 
\begin{enumerate}
	\item The causal effect of any variable (say, $D$) on another variable (say, $Y$) is defined by a nonlinear function.
	\item The effects of unobserved variables on $Y$ can be expressed by a single error term that is independent of $D$ and does not interact with $D$ (such that the effects of unobserved variables on $Y$ are homogeneous across values of $D$). 
\end{enumerate}	
In other words, if causal associations between observed variables are nonlinear and causal effects of unobserved variables can be represented by additively separable (i.e., non-interacting) and independent error terms, then we can deduce the direction of causality between observed variables from the data. Formally, this requires that $Y$ is characterized by the following model: 
\begin{eqnarray}\label{nonlinear} 
Y=\mu(D)+\varepsilon,
\end{eqnarray}
where $\mu$ is a nonlinear function of $D$ and $\varepsilon$ is an additive error term that is independent of $D$. 

The nonlinearity of $\mu$ carries an important implication. When erroneously assuming that $Y$ affects $D$ and estimating a reverse (or ``anticausal'') association of $D$ as a function of $Y$, the resulting error term will not be independent of $Y$. This holds true except for highly specific cases, as elaborated in \cite*{peters2017elements}. Notably, in scenarios beyond these special cases, this phenomenon enables the identification of the correct causal model, wherein $D$ affects $Y$. In this correct model, the error term is independent of $D$. In practice, testing the causal direction may be based on running nonlinear regressions of both $Y$ on $D$ and $D$ on $Y$ and verifying in which of these two cases the estimated errors (or residuals) are independent of the regressors. To this end, we can apply a statistical test for the independence of the estimated errors and regressors in either regression and choose that causal model as the supposedly correct one for which the test yields a higher (and statistically insignificant) p-value. To avoid overfitting bias when computing the p-values, sample splitting is advisable, which implies estimating the regression models and conducting the independence tests in distinct subsamples obtained from randomly splitting the full sample. Sample splitting ensures that the regression and testing stages are not statistically associated with each other. 

The model in equation \eqref{nonlinear} can be extended or modified in various ways, all while preserving the ability to identify causality. For instance, \cite*{breunig2021testability} consider testing the direction of causality between two variables $D$ and $Y$  when controlling for observed covariates, denoted by $X$, while allowing for heteroskedasticity of $\varepsilon$ in $X$. As the independence between $\varepsilon$ and $D$ is only required to hold conditional on $X$, this implies a type of selection-on-observables assumption. Formally, the setup is defined by the following nonlinear model:
\begin{eqnarray}\label{nonlinear3}
Y = \mu(D, X) + \varepsilon,
\end{eqnarray}
where the variance of error term $\varepsilon$ may differ across values of $X$. 

Furthermore,  \cite*{zhang2006extensions} and \cite*{zhang2009identifiability} consider a generalization of equation \eqref{nonlinear}, known as post-nonlinear (PNL) model, which consists of two nested non-linear functions: 
\begin{eqnarray}\label{nonlinear2} 
	Y=q(\mu(D)+\varepsilon),
\end{eqnarray}
where both $q$ and $\mu$ are non-linear functions and $q$ is assumed to be invertible. It is evident that equation \eqref{nonlinear} is a special case of equation \eqref{nonlinear2} when $q$ is defined as the identity function. For this reason, equation \eqref{nonlinear2} is more general, because it permits $Y$ to be a complex (rather than additive) function of the impact of the treatment, characterized by $\mu(D)$, and of unobserved variables, characterized by $\varepsilon$. Another special case of both equations \eqref{nonlinear} and  \eqref{nonlinear2} is a model in which the causal associations between observed variables $D$ and $Y$ are linear rather than nonlinear:
\begin{eqnarray}\label{linearlearning} 
	Y=\alpha+\beta D+\varepsilon,
\end{eqnarray}
where $\alpha$ is the constant term and $\beta$ is the causal effect of $D$ on $Y$. \cite*{shimizu2006linear} show that the correct causal model is identified if the additively separable error term $\varepsilon$ is non-normally distributed. The latter assumption is crucial, as unique identification in linear models is generally not feasible if errors are normally distributed, unless additional assumptions are imposed. For instance, \cite*{PetersB2013} show that in linear models with normally distributed errors, causality can still be learned under the (strong) assumption that the error terms in the various equations of a causal model all have the same variance. 

%Learning which variables\enlargethispage{-1\baselineskip} affect which other variables from statistical associations alone rather than from a presupposed causal structure, in which a treatment affects the outcome but not vice versa, is a challenging task. However, a growing number of studies demonstrates under which assumptions and circumstances this is at least theoretically feasible; for instance, see \cite*{KalischB2014}; \cite*{peters2017elements}; \cite*{Glymouretal2019}. Should causal discovery succeed in correctly revealing causal associations on a larger scale in practically relevant cases, then it would arguably be the first artificial intelligence method that comes closer to actually doing something intelligent: finding complex causal associations, which are a fundamental part of human reasoning. Therefore, the best of causal analysis might be yet to come, so let's stay tuned for exciting developments in the future.

We subsequently illustrate the implementation of causal discovery algorithms with a simple empirical example using the statistical software \textsf{R}. To this end, the following discussion will present some syntax in \textsf{R}, which can be skipped by readers who are not interested in this detail. We first install the \textsf{R} packages  \textit{bnlearn}  and \textit{causalweight} created by \cite*{Scutari2010} and \cite*{BodoryHuber2018}, respectively, using the command \textit{install.packages("bnlearn", "causalweight")}. We then load both packages using the \textit{library} command. Next, we utilize the \textit{data} command to load the \textit{coffeeleaflet} data set, which stems from an experimental study aimed at evaluating the impact of a leaflet discussing coffee production implications on the environmental awareness of students in Bulgaria. 

\begin{figure}[!htp]
	\centering \caption{\label{colliderbias}  Collider structure}
	\begin{center}
		\centering 	\includegraphics[scale=0.5]{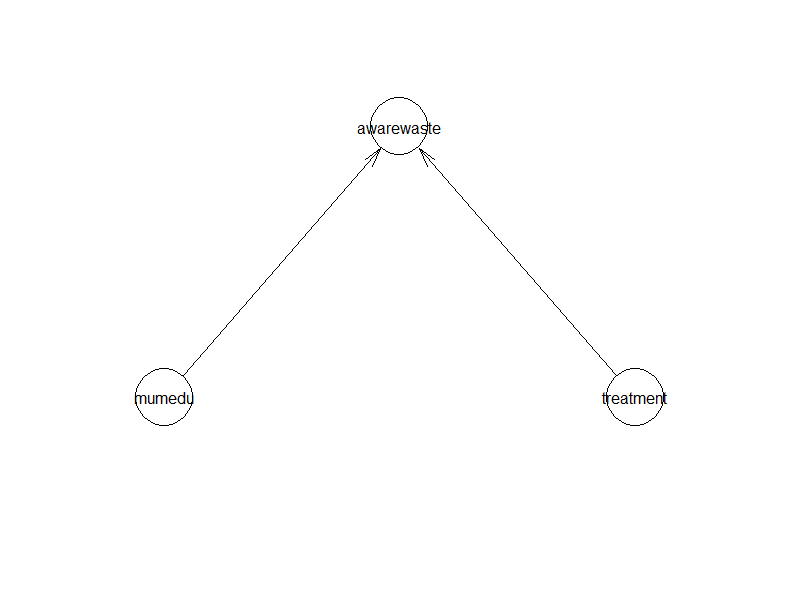}
	\end{center}
\end{figure}

We create a new frame set named \textit{data}, which only contains the pretreatment characteristic ``mother's education'' (the variable \textit{mumedu} in the 8th column of the \textit{coffeeleaflet} data), the randomly assigned leaflet treatment (the variable \textit{treatment} in the 32nd column), and the outcome ``awareness of waste production due to coffee production'' measured on a five-point scale (the variable \textit{awarewaste} in the 38th column): \textit{data=coffeeleaflet[,c(8,32,38)]}. As the variable \textit{treatment} is randomly assigned, it should be independent of \textit{mumedu}, while both variables may affect the outcome \textit{awarewaste}. In this case, we have the causal relation \textit{treatment}$\rightarrow$\textit{awarewaste}$\leftarrow$\textit{mumedu}, such that \textit{awarewaste} is a collider. We feed our data set into the \textit{pc.stable} function to conduct causal discovery based on the PC algorithm and store the results in an \textsf{R} object named \textit{output}. Finally, we employ the \textit{plot} command to visualize the causal graph derived from the PC algorithm. The box below provides the \textsf{R} code for each of the steps.
\begin{tcolorbox}
	\begin{lstlisting}
install.packages("bnlearn", "causalweight") # install packages
library(bnlearn)                            # load bnlearn package
library(causalweight)                       # load causalweight package
data(coffeeleaflet)                         # load coffeeleaflet data
data=coffeeleaflet[,c(8,32,38)]             # select variables nr. 8, 32, and 38
output=pc.stable(data)                      # run the PC algorithm
plot(output)                                # plot the learned causal model
	\end{lstlisting}
\end{tcolorbox}
Executing the code produces the DAG depicted in figure \ref{colliderbias}. We see that the algorithm detects the previously mentioned collider structure, implying that mother's education and treatment assignment are independent (as expected in a well conducted experiment), while they both affect the awareness outcome.  As mentioned before, such algorithms might fail to uniquely determine the direction of the causal arrows in more complicated models with more variables and more involved dependence structures than in our toy model containing a single collider. And even in our toy model, we acknowledge that the causal arrow from mother's education to the awareness outcome might be confounded, as mother's education is not randomly assigned.

As a further \textsf{R} example, we consider learning the direction of causality in nonlinear models characterized by equation \eqref{nonlinear3} by testing the independence of the estimated nonlinear regression function and the error terms (or residuals). Assuming that all required \textsf{R} packages have been previously installed (using the \textit{install.packages} command), we apply the \textit{library} command to load the \textit{dHSIC} package for independence testing, the \textit{mgcv} package for running nonlinear regressions, and the \textit{datarium} package. The latter contains the \textit{marketing} data set consisting of 200 observations with information on sales and advertising budgets, which we load using the \textit{data} command. We are interested in learning the causal relationship between the variable \textit{newspaper}, which measures the budget of advertisement in newspapers and which we define as treatment \textit{D}, and the variable \textit{sales}, which we define as outcome \textit{Y}. 

We set a seed (\textit{set.seed(1)}) for the replicability of our results to follow and use the \textit{gam} function to estimate two different regression models. In the first regression, we estimate the outcome \textit{Y} as a nonlinear function of the treatment \textit{D}: \textit{model1 = gam(Y $\sim$ s(D))}, where the wrapper \textit{s(D)} implies that the association is measured by series regression. We store the output in an \textsf{R} object named \textit{model1}. In the second (reverse) regression, we estimate the treatment \textit{D} as a nonlinear function of the outcome \textit{Y} and store the output in the object \textit{model2}. We then feed the residuals of the first regression model, \textit{model1\$residuals}, and \textit{D} into the \textit{dhsic.test} command to test the independence between both variables and save the results in object \textit{test1}. Using the same command, we test the independence between the residuals of the second regression model and \textit{Y} and save the results in object \textit{test2}. Finally, we investigate the p-values of the tests stored in \textit{test1\$p.value} and \textit{test2\$p.value}.
\begin{tcolorbox}
	\begin{lstlisting}
library(dHSIC)                        # load dHSIC package
library(mgcv)                         # load mgcv package
library(datarium)                     # load datarium package
data(marketing)                       # load marketing data
D=marketing$newspaper                 # define treatment (newspaper advertising)
Y=marketing$sales                     # define outcome (sales)
model1=gam(Y ~ s(D))                  # estimate Y as nonlinear function of D
model2=gam(D ~ s(Y))                  # estimate D as nonlinear function of Y
set.seed(1)                           # set seed
test1=dhsic.test(model1$residuals, D) # independence test for first model
test2=dhsic.test(model2$residuals, Y) # independence test for second model
test1$p.value; test2$p.value          # show p-values
	\end{lstlisting}
\end{tcolorbox}
The tests yield p-values of 0.136 (or roughly 14\%) and 0.012 (or roughly 1\%) for the regressions of \textit{Y} on \textit{D} and of \textit{D} on \textit{Y}, respectively. Consequently, we find that a causal effect of the supposed treatment \textit{D} on outcome \textit{Y} cannot be rejected at the 10\% level of statistical significance. In contrast, a reverse causal effect from \textit{Y} to \textit{D} is rejected at the 5\% level of significance.

\section{The Back-Door and Front-Door Criteria}\label{backfront}

A concept closely related to d-separation discussed in section \ref{dsep} is the back-door criterion, see e.g.\ \cite*{Pearl00}. It provides a causal graph-based framework for determining variables that must be controlled for to prevent confounding bias when measuring the causal effect of a treatment variable $D$ (like education), on an outcome variable $Y$ (like earnings). More concisely, the back-door criterion employs d-separation to detect an appropriate set of control variables such that any confounding association between $D$ and $Y$ (e.g. through characteristics like socio-economic status), also known as back-door path, is eliminated or blocked. More formally, a set of covariates $X$ satisfies the back-door criterion in a DAG where the effect of $D$ on $Y$ is of interest, 
\begin{enumerate}
	\item if $X$ blocks any path between $D$ and $Y$ that contains an arrow (causal effect) into $D$,
	\item if no variable in $X$ is causally affected by $D$.
\end{enumerate}

To provide an example illustrating the concept of a back-door path and the back-door criterion, let us reconsider the left graph in figure \ref{DAGs}. In this case, the presence of covariates $X$ (like socio-economic status) that jointly affect $D$ (education) and $Y$ (earnings) forms a back-door path ($D\leftarrow X \rightarrow Y$) which confounds the causal relation of $D$ and $Y$. Controlling for $X$ satisfies the back-door criterion, as it closes the back-door path (i.e.\ controls for confounding), while not including any variable that is causally affected by $D$. This implies that the so-called selection-on-observables or unconfoundedness assumption, which is frequently invoked in policy or treatment evaluation studies, is satisfied with respect to $D$, meaning that the treatment is as good as randomly assigned conditional on $X$. Consequently, the treatment effect of $D$ on $Y$ is identified when controlling for $X$. For a binary treatment $D$, the average treatment effect (ATE) of providing everyone vs.\ no-one with the treatment, henceforth denoted by $\Delta$, is for instance identified by the following expression: 
\begin{eqnarray}\label{backdoorate}
	\Delta=E[E[Y|X,D=1]-E[Y|X,D=0]].
\end{eqnarray} 
See for instance \cite*{Im04}, \cite*{ImWo08}, and \cite*{AbadieCattaneo2018} for a discussion of the unconfoundedness assumption and evaluation methods for ATE estimation like regression, matching, weighting, or combinations thereof like doubly robust estimation. 

\tikzstyle{EdgeStyle}   = [->,>=stealth']
\begin{figure}[!htp]
	\centering \caption{\label{DAGs2}  M-bias}\bigskip
	\begin{tikzpicture}[scale=1.3]
		\SetGraphUnit{2}
		\Vertex{D}    \NOEA(D){X} \NOWE(X){V} \NOEA(X){U}  \SOEA(X){Y}
		\Edges(D,Y) \Edges(V,D) \Edges(U,Y) \Edges(V,X) \Edges(U,X)
	\end{tikzpicture}
\end{figure}

In contrast to the previous example, figure \ref{DAGs2} presents a causal model where controlling for $X$ does not satisfy the back-door criterion. Although the back-door condition that $X$ is not causally affected by $D$ holds, the condition that $X$ blocks any path between $D$ and $Y$ containing an arrow into $D$ is not met. This situation arises because $X$ is affected by both $V$ and $U$ and thus a collider, such that controlling for $X$ introduces dependence between $V$ and $U$. As $V$ affects $D$ and $U$ affects $Y$, this collider bias generates a dependence or back-door path between $D$ and $Y$, leading to the failure of the back-door criterion. This specific form of collider bias is known as M-bias. However, the back-door criterion holds when not controlling for $X$, thus preventing collider bias, as $V$ (which influences $D$) is not associated with $Y$ when $X$ is not controlled for. Alternatively, the back-door criterion can be met by controlling for $X$ and either $V$ or $U$, or both. The rationale here is that controlling for either $V$ or $U$ blocks the confounding bias between $D$ and $Y$ that is introduced by controlling for $X$.

It is worth noting that a web-based application called ``DAGitty'' (Directed Acyclic Graph Interactive Tool) for creating, analyzing, and visualizing causal graphs is available at \href{https://dagitty.net/}{https://dagitty.net/}. It makes use of the back-door criterion to  determine if and under which conditions the effect of a treatment $D$ on an outcome $Y$ is identified in a causal model. DAGitty provides a visual representation of causal graphs, allows differentiation between observed and unobserved variables, and highlights all back-door paths between $D$ and $Y$ that need to be controlled for even in complex models with numerous variables. This tool proves useful for researchers and analysts in comprehending the underlying causal structure and making correct decisions about the feasibility of identifying treatment effects based on observed variables. If identification is possible, DAGitty also indicates the covariates that need to be controlled for. DAGitty is also available as \textsf{R} package, which is provided by \cite*{Textoretal2017}.

A further criterion for identifying the causal effect of $D$ on $Y$ is the so-called front-door criterion, see \cite*{Pearl00}, which does not rely on blocking all back-door paths between $D$ and $Y$. Yet, it builds on the back-door criterion by applying it sequentially with respect to the effect of $D$ (like education) on a mediating variable $M$ (like human capital acquisition), through which presumably any effect of $D$ on $Y$ (like earnings) operates, and the effect of $M$ on $Y$. Formally, a set of variables $M$ satisfies the front-door criterion for identifying the causal effect of $D$ on $Y$, 
\begin{enumerate}
	\item if any effect of $D$ on $Y$ operates via $M$,
	\item if there is no unblocked back-door path from $D$ to $M$,
	\item if all back-door paths from $M$ to $Y$ are blocked when controlling for $D$. 
\end{enumerate}
The first assumption states that $M$ fully mediates the effect of $D$ on $Y$. The second assumption imposes that no variables jointly affect $D$ and $M$, ensuring that the treatment-mediator relation is unconfounded. The third assumption imposes that conditional on $D$, no variables jointly affect $M$ and $Y$, such that the mediator-outcome relation is unconfounded conditional on the treatment. Alongside these assumptions, the identification of causal effects based on the front-door criterion requires a specific common support restriction: For a binary treatment $D$, it must hold that $0<\Pr(D=1|M)<1$, ensuring that both treated and untreated observations exist for each possible value of $M$ in the population. 

\tikzstyle{EdgeStyle}   = [->,>=stealth']
\begin{figure}[!htp]
	\centering \caption{\label{frontdoorfig}  Front-door criterion without and with controlling for covariates (left and right)}\bigskip
	\begin{tikzpicture}[scale=1.3]
		\SetGraphUnit{2}
		\Vertex{D}  \EA(D){M}  \EA(M){Y} \SO(M){U}
		\Edges(D,M,Y)  
		\draw[dashed,->] (U) to (D);
		\draw[dashed,->] (U) to (Y);
	\end{tikzpicture}$\hspace{6em}$
	\begin{tikzpicture}[scale=1.3]
		\SetGraphUnit{2}
		\Vertex{D}  \EA(D){M}  \EA(M){Y} \SO(M){U} \NO(M){X}
	\Edges(D,M,Y)  	
	\Edges(X,M) \Edges(X,D) \Edges(X,Y)
	\draw[dashed,->] (U) to (D);
	\draw[dashed,->] (U) to (Y); 
	\end{tikzpicture}
\end{figure}
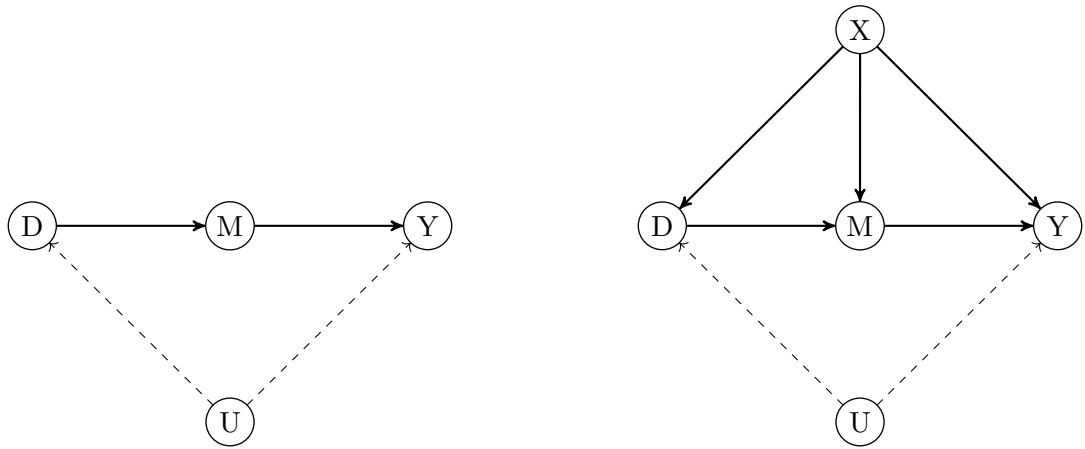 

The front-door criterion is related to both causal mediation models and instrumental variables, see for instance \cite*{Huber2019} and \cite*{HuberWuthrich2019} for literature surveys. The assumptions imply ``complete'' mediation in the sense that there only exists an indirect effect of $D$ on $Y$ via $M$ but no direct effect, similar to the exclusion restriction in instrumental variable methods. The left graph in figure \ref{frontdoorfig} provides an illustration of the front-door criterion. $U$ denotes unobserved confounders of $D$ and $Y$, where the dashed arrows imply that the causal paths from $U$ cannot be observed. In contrast, confounders of the causal association between treatment $D$ and mediator $X$ (which is on the causal path between $D$ and $Y$) or between mediator $M$ and outcome $Y$ conditional on $D$ must not exist. This still permits unobservables affecting $M$ (which are omitted in the graph), as long as those unobservables do not affect $D$ or $Y$, too, and are not associated with $U$. If the front-door criterion holds, the ATE corresponds to the following expression, see for instance \cite*{frsp2019}:
\begin{eqnarray}\label{frontdoorate}
	\Delta=E[ \nu(M) |D=1]-E[ \nu(M) |D=0].
\end{eqnarray}
where $\nu(m)=E[E[Y|D,M=m]]$ is a nested mean outcome, in which the conditional mean outcome $E[Y|D,M=m]$ is averaged over values of the treatment $D$, while keeping the mediator fixed at value $M=m$. For a binary treatment, it holds that  
$\nu(m)=E[Y|D=1,M=m]\cdot \Pr(D=1)+E[Y|D=0,M=m]\cdot \Pr(D=0)$, with $\Pr$ denoting a probability. 

Equation \eqref{frontdoorate} calculates ATE as the average change in outcome $Y$ due to a shift in mediator $M$ induced by switching the treatment from $D=1$ to $D=0$, which utilizes the first assumption that any effect of $D$ on $Y$ operates through $M$. Furthermore, the third assumption ensures that $D$ blocks all back-door (or confounding) paths from $M$ to $Y$, implying that by controlling for $D$ in the nested mean outcome $\nu(M)$, we can assess the effect of $M$ on $Y$ by varying the values of $M$ in $\nu(M)$. Finally, given the second assumption that there is no back-door path between $D$ and $M$, the effect of $D$ on $M$ is identified as well. As a result, we can compute the mean potential outcomes and thus, the ATE by simply averaging $\nu(M)$ within the treated and nontreated groups.

%While equation  \eqref{frontdoor} identifies mean potential outcomes as required for the evaluation of the ATE, the front-door criterion was originally formulated in terms of the probability that a potential outcome takes a specific value $y$, denoted as $\Pr(Y(d)=y)$. As discussed in \cite*{Pearl95}, this potential outcome probability is identified through the following equation:
%\begin{eqnarray}
	%\Pr(Y(d)=y)=\sum_m \Pr(M=m|D=d)\cdot \left(\sum_{d'} \Pr(Y=y|D=d',M=m)\cdot \Pr(D=d')\right), %\notag\\
	%\Pr(Y(0)=y)&=&\sum_m \Pr(M=m|D=0)\sum_d \Pr(Y=y|D=d,M=m)\Pr(D=d),
%\end{eqnarray}
%where we assume that all of $Y,D,M$ are discrete variables such that we can measure their probabilities. 

The front-door criterion has been rarely applied in empirical research and is virtually absent in social sciences. One potential application discussed in \cite*{Pearl00} is the assessment of the causal effect of smoking ($D$) on lung cancer ($Y$) mediated by tar deposits ($M$). If unobserved genetic factors ($U$) simultaneously influence smoking behavior and lung cancer, then the back-door criterion fails for $D$ and $Y$ due to confounding. Nevertheless, the effect of smoking on lung cancer can potentially be identified if the front-door criterion is met. This criterion necessitates that there are no unobservable variables affecting both smoking behavior and tar deposits, or tar deposits and lung cancer, and that smoking solely affects lung cancer through tar deposits (exclusion restriction). As noted e.g.\ by \cite*{imbens2020potential}, the restrictions on the unobservables imposed by the front-door criterion fail in this application if genetic factors influence smoking behavior, the inclination to accumulate tar, and lung cancer altogether, or if unobservables (like a hazardous working environment) jointly affect tar deposits and lung cancer. Additionally, if there exist alternative causal pathways through which smoking behavior affects lung cancer beyond tar deposits, then the exclusion restriction assumption fails. In numerous empirical contexts, meeting the requirements of the front-door criterion appears challenging, potentially explaining its limited prevalence in practical applications. 

However, it is worth mentioning that the assumptions underlying the front-door criterion can be relaxed to only hold conditionally on observed variables. Indeed, the assumptions might appear more plausible in empirical applications when allowing covariates $X$ to affect all of $D$, $M$, and $Y$, as illustrated in the right graph of figure  \ref{frontdoorfig}. In this case, the ATE is identified when controlling for $X$. Consequently, this entails modifying expression \eqref{frontdoorate} to 
\begin{eqnarray}\label{frontdoor2}
	\Delta= E_X[E_M[ \nu(M,X) |D=1,X]-E_M[ \nu(M,X) |D=0,X]],
\end{eqnarray}
where $\nu(m,x)=E[E[Y|D,M=m, X=x]|X=x]$ is a nested conditional mean outcome given $X$. Intuitively, one performs similar calculations for the previous case without covariates, but this time within groups that share the same values of the covariates $X$ to obtain the conditional ATE (CATE), assuming that the front-door criterion holds conditional on $X$. Finally, one averages the CATE across all values of $X$ (as indicated by the subscript $X$ in the expectation operator $E_X$) to obtain the ATE for the entire population. 

%Another relaxation of the front-door criterion is to allow for a direct effect of $D$ on $Y$ that does not operate through $M$, thereby violating the exclusion restriction. As demonstrated in \cite*{Fulcheretal2019}, it is then still feasible to assess the indirect treatment effect of $D$ on $Y$ that operates via $M$, whereas the (total) ATE of $D$ on $Y$ and the direct effect (not operating via $M$) are not recovered. Furthermore, the authors propose a doubly robust (DR) method, which provides an estimate of the indirect effect when the exclusion restriction is violated (but the other front-door assumptions hold) and an estimate of the ATE when all front-door assumptions hold, see also the discussion in \cite*{Gorbachetal2023}. Yet another DR estimator permits assessing causal effects when the front-door criterion is satisfied conditional on covariates $X$, as considered in equation \eqref{frontdoor2}. This estimator may be combined with machine learning to control for covariates in a data-driven way. 

\section{Examples of Causal Discovery}\label{examples}

As discussed in section \ref{causaldisalg}, attempting to uncover all causal relationships between all observed (or even unobserved) variables in a causal model can be very challenging or practically infeasible in many empirical scenarios. This section delves into a few examples where causal discovery is applied to somewhat more modest objectives than learning an entire causal structure. Yet, these objectives may appear interesting from the perspective of economics or social sciences and can be more feasible from a practical standpoint.

For example, researchers or analysts might have an interest in identifying all observed variables that directly influence a specific outcome variable $Y$. In other words, one aims at detecting all treatments exerting a direct causal impact on $Y$. For instance, we may want to determine all observed variables that have an effect on earnings (which could include education, profession, work experience, labor market conditions, among others) or affect health (which could include health behavior, genetic factors, living/working environment, among others). Learning treatments requires imposing specific identifying assumptions, such as the condition that any variables jointly affecting any treatment and the outcome are observed in the data. This implies that the selection-on-observables assumption holds for each of the potentially numerous treatments under consideration, which is a strong condition. Furthermore, we need to rule out reverse causal effects of the outcome $Y$ on the treatments. This appears plausible if the observed variables are measured at an earlier point in time than the outcome, as variables measured later cannot impact the values of variables measured earlier.

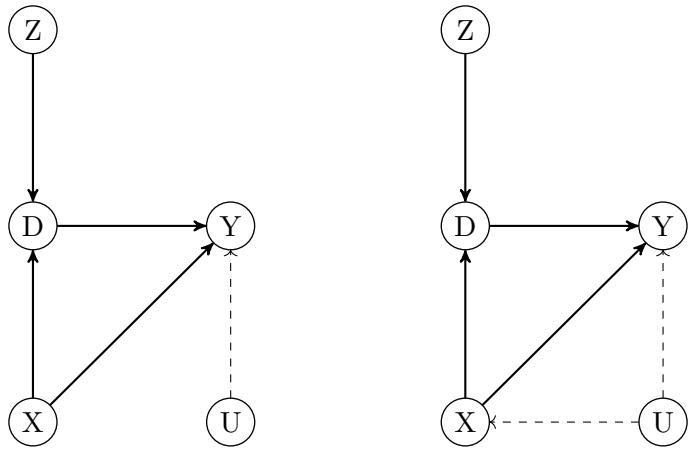
\begin{figure}[!htp]
	\centering \caption{\label{DAGexamples}  Learning treatments and/or the identification of treatment effects}\bigskip
	\begin{tikzpicture}[scale=1.3]
				\SetGraphUnit{2}
		\Vertex{D}  \EA(D){Y}  \SO(D){X} \NO(D){Z} \SO(Y){U}
		\Edges(D,Y) \Edges(X,D) \Edges(X,Y) \Edges(Z,D)
		\draw[dashed,->] (U) to (Y);
	\end{tikzpicture}$\hspace{6em}$
\begin{tikzpicture}[scale=1.3]
\SetGraphUnit{2}
	\Vertex{D}  \EA(D){Y}  \SO(D){X} \NO(D){Z} \SO(Y){U}
\Edges(D,Y) \Edges(X,D) \Edges(X,Y) \Edges(Z,D)
\draw[dashed,->] (U) to (Y);
\draw[dashed,->] (U) to (X); 
\end{tikzpicture}
\end{figure}

The graph on the left in figure \ref{DAGexamples}  provides a visual illustration, in which the observed variables $D$ and $X$ causally affect the outcome $Y$. In contrast, $Z$ is not part of the treatments directly affecting $Y$, as it only exerts an indirect effect on the outcome through $D$. For both $D$ and $X$, the selection-on-observables assumption holds in the left graph, as the unobserved term $U$ affecting the outcome (where the non-observability of the effect is indicated by the dashed line) neither affects $D$ nor $X$. Conversely, the selection-on-observables assumption fails for $X$ in the right graph of figure \ref{DAGexamples}, where $U$ jointly affects $X$ and $Y$. 

\tikzstyle{EdgeStyle}   = [->,>=stealth']

\cite*{soleymani2022causal} provide an algorithm in this context that selects treatment variables in a data-driven way, while also controlling for observed covariates in a data-adaptive manner based on the double machine learning (DML) framework, a causal machine learning approach suggested by \cite*{Chetal2018}. The algorithm sequentially considers each of the observed variables that may affect $Y$ as treatment $D$ to estimate the direct effect of that candidate treatment on the outcome $Y$, while considering all remaining observed variables as covariates $X$ to be controlled for by DML. Finally, the algorithm retains only those observed variables as treatment variables that exhibit statistically significant effects on $Y$ (where judging statistical significance should account for issues related to testing multiple hypotheses based on multiple candidate treatments). Another machine learning-based algorithm proposed by \cite*{quinzan2023drcfs} adopts a somewhat distinct testing approach. In a first step, the algorithm estimates the outcome as a function of all observed variables by DML. In a second step, it repeats the estimation of the outcome when excluding one candidate treatment $D$. If the omission of $D$ statistically significantly alters the outcome estimate compared to using all observed variables (as done in the first step), then $D$ is retained as treatment variable, otherwise it is dropped. By sequentially applying this process to all candidate treatments, the algorithm provides an estimated set of treatments that have a direct effect on $Y$. 

Another example of applying causal discovery concepts involves learning whether sufficient conditions for identifying the causal effect of a treatment of interest ($D$) on an outcome of interest ($Y$) hold in observational data when imposing causal faithfulness (see section \ref{dsep}). In contrast to the previous example, the objective here is not to learn all treatments influencing an outcome, but rather to assess whether the effect of a predefined treatment is identifiable within the available data. As discussed in \cite*{deLunaJohansson2012}, \cite*{BlackJooLaLondeSmithTaylor2015}, and \cite*{huber2023testing}, if there exists an instrument $Z$ that affects $D$ and is conditionally independent of $Y$ given $X$, then this is a sufficient condition for the selection-on-observables assumption. Consequently, the causal effect of $D$ on $Y$ is identified conditional on $X$. This result follows from an application of the d-separation criterion: If covariates $X$ permit controlling for all back-door paths (or confounders) between $Z$ and $Y$ as well as $D$ and $Y$ and if $Z$ is a valid instrument in the sense that it does not directly affect the outcome $Y$ other than through $D$, then controlling for $D$ in addition to $X$ leads to the independence of $Z$ and $Y$. In particular, conditioning on $D$ does not introduce collider bias (see section \ref{dsep}) between $Z$ and $Y$ given $X$, implying that $D$ is as good as randomly assigned conditional on $X$, such that the selection-on-observables assumption holds. In both graphs of figure \ref{DAGexamples}, the testable implication holds for treatment $D$, implying that the effect of $D$ on $Y$ is identified conditional on $X$. Instrument $Z$ affects outcome $Y$ only through treatment $D$ and the unobservable $U$ does not jointly affect $Z$ and $Y$ or $D$ and $Y$ conditional on $X$. Therefore, $Z$ and $Y$ are conditionally independent given $X$ and $D$. 

As an example, consider the case that conditional on observed socio-economic characteristics $X$, there are no unobserved factors jointly affecting participation in a training program ($D$) and earnings ($Y$), such that selection-on-observables holds. Furthermore, assume there is a random encouragement or invitation to attend the program ($Z$) , which induces some of the (randomly chosen) invitees to attend the training, but has no direct effect on earnings, such that $Z$ is an instrument for $D$. In this setup satisfying selection-on-observables and instrument validity, it holds that the encouragement is statistically independent of the earnings outcome conditional on training participation and the socio-economic characteristics.
It is interesting to note that this approach can also be applied in the context of comparing treatment effects from experimental and observational data, as seen in the study by \cite*{morucci2023double}. In this context, \( Z \) is a binary indicator for experimental or observational data, while \( D \), \( X \), and \( Y \) represent the treatment, covariates, and outcome measurements in the respective datasets. The conditional independence of \( Z \) and \( Y \) given \( D \) and \( X \) in the joint data again implies the satisfaction of both the selection-on-observables assumption with respect to \( D \) and the instrument validity of \( Z \). However, the latter assumption now bears a specific interpretation: Conditional on \( X \), treatment effects are homogeneous across the experimental and observational data; otherwise, \( Z \) would have a direct association with \( Y \) through effect heterogeneity. Such heterogeneous effects could, for example, stem from differences in the timing or geographic location in which the experimental and observational data were collected, meaning that the treatment (like a training program) is more effective in some regions or periods (e.g. with a particularly high or low unemployment rate) than in others.

The previous insights on the implications of conditional independence can be used in a causal discovery approach to learn sets of instruments $Z$ and covariates $X$ that satisfy the conditional independence of $Z$ and $Y$ given $X$ and $D$ in the data (if such sets exist at all), as outlined in \cite*{Apfel2023learning}, rather than predefining $Z$ and $X$. The method is based on considering all observed variables which are not the treatment $D$ or the outcome $Y$ and assuming one of them to be a candidate instrument $Z$, while the remaining ones (or subsets thereof) constitute the covariates $X$. In a first step, the first-stage effect of $Z$ on $D$ given $X$ is estimated by DML to verify whether the conditional association between $Z$ and $D$ is strong enough, which is a precondition for testing. Only if this is the case, the second step consists of testing the conditional independence of $Z$ and $Y$ given $X$ and $D$ by DML. This two-step approach is iteratively repeated to each variable that is neither $D$ nor $Y$, effectively cycling through all candidates for the role of the instrument $Z$. 

For instance, assume that $D$ is college education and $Y$ is later life earnings, while the remaining variables are socio-economic characteristics like parental education, parental income, and place of residence, among others. The algorithm would initially designate parental education as $Z$ and the remaining variables (or their subsets) as $X$, followed by treating place of residence as $Z$ with the remaining variables as $X$, and so forth. If for one or even several definitions of $Z$ and $X$ the conditional independence assumption is not rejected by the conditional independence tests, this signals the possible fulfillment of the selection-on-observables assumption, particularly in large samples with a reduced uncertainty in testing. Consequently, one can opt for the definition of control variables $X$ (and $Z$) that yields the highest p-value when running the conditional independence test, thereby implying the lowest probability of violating conditional independence. Relatedly, \cite*{hassanpour2019learning} and \cite*{wu2021learning} propose so-called deep learning algorithms to simultaneously (rather than iteratively) learn sets of variables which are (1) observed confounders $X$ that jointly affect a predefined treatment $D$ and outcome $Y$, (2) instruments $Z$ that only affect the treatment $D$, and (3) outcome predictors that only affect the outcome $Y$.

Another causal discovery method that combines elements of the previous examples is suggested in \cite*{peters2015causal}. This approach aims at the identification of multiple treatments (like in the first example), which directly affect an outcome $Y$ and satisfy a selection-on-observables assumption, by means of instruments (like in the second example), also referred to as ``environments'' in the computer science literature. If an instrument $Z$ affects outcome $Y$ only through one or several other observed variables and is independent of $Y$ conditional on a set of candidate treatment variables (rather than a single treatment), then this set will include variables that have a direct causal effect on $Y$. Another way of putting this is that if e.g.\ the conditional mean outcome does not depend on (or is invariant across) variations in instrument $Z$ conditional on a set of candidate treatments $\tilde{D}$, such that $E[Y|\tilde{D},Z]=E[Y|\tilde{D}]$, then $\tilde{D}$ contains variables affecting $Y$ because this set intercepts any (average) effect of $Z$ on $Y$.  In light of this, \cite*{peters2015causal} propose an algorithm that aims at finding sets of $\tilde{D}$ for which the estimation of  $E[Y|\tilde{D}]$ remains invariant
 across different values of $Z$. The method's capability of detecting treatments directly affecting the outcome hinges on the strength of the association between $Z$ and the treatments. If the first-stage association is weak or absent, the estimation of $E[Y|\tilde{D}]$ will be (almost) invariant across values of $Z$ even if important treatments are missing in set $\tilde{D}$. For this reason, more and stronger instruments increase the chance of finding true treatments that directly affect $Y$.

A final example concerns testing the identification of the causal effect of $D$ on $Y$ (as in the second example) when substituting instrumental variable assumptions on $Z$, which rule out a direct effect of $Z$ on $Y$ or an association between $Z$ and unobservables $U$ affecting $Y$, with a different set of assumptions. \cite*{karlsson2023detecting} present a method under the assumption that $Z$, representing specific environments like distinct observational datasets, directly influences all of the covariates $X$, treatment $D$, outcome $Y$, and unobservables $U$ (which impact $Y$). However, the mechanisms through which $Z$ affects each of $X$, $D$, $U$, and $Y$ must be independent of one another. In other words, $Z$ is presumably characterized by a set of variables that are mutually independent, with each variable only influencing one of $X$, $D$, $U$, and $Y$. When represented as $Z=(Z_X,Z_D,Z_U,Z_Y)$, the assumption of independent causal mechanisms implies, for example, that the effect of $Z_D$ on $D$ must not be associated with the effect of $Z_U$ on $U$.

Under these conditions, the selection-on-observables assumption for treatment $D$ can be tested using the following algorithm. First, an observation with index $i$ is randomly selected (e.g. $i=1$ for the first randomly sampled observation) in each environment $z \in {1,2,...,\mathcal{Z}}$, where $\mathcal{Z}$ is the number of different environments. Denoting $D_i(z)$ as the treatment state of the observation with index $i$ in environment $z$, the next step involves aggregating the treatments of index $i$ across all environments into a treatment vector, denoted as $\mathcal{D}_i = (D_i(1),D_i(2),...,D_i(\mathcal{Z}))$. Proceeding analogously for covariates $X$ and outcome $Y$, the vectors $\mathcal{X}_i$ and $\mathcal{Y}_i$ are obtained. The final step is to test whether, for any distinct indices $i\neq j$ (e.g., $i=1$ and $j=2$), $\mathcal{D}_j$ is conditionally independent of $\mathcal{Y}_i$ when controlling for $\mathcal{D}_i$, $\mathcal{X}_i$, and $\mathcal{X}_j$. A rejection of conditional independence indicates a violation of the selection-on-observables assumption. It is worth noting that the choice of observations assigned indices $i$ and $j$ within an environment is arbitrary if the samples in each environment are drawn randomly; the only requirement is that $i$ and $j$ cannot be assigned to the same observation within an environment.

As an empirical example in \textsf{R}, let us consider testing whether the causal effect of a predefined treatment is identified in observational data based on a predefined instrument, using the method of \cite*{huber2023testing}. We load the \textit{causalweight} package and use the \textit{data} command to load the \textit{JC} data from an experimental study of the Job Corps program, a training program for disadvantaged youth in the US, see \cite*{ScBuMc2008}. The goal is to test whether the average effect of training participation in the first year after program assignment ($D$) on the health state four years after assignment ($Y$) is identified when controlling for the baseline covariates ($X$) measured prior to training. To this end, we test whether random assignment to Job Corps ($Z$), which has a strong first-stage effect on $D$, is conditionally mean independent of $Y$ given $D$ and $X$. To perform this test, we define the variables \textit{Z=JC[,1]} (as the first column in the \textit{JC} data contains the binary assignment variable), \textit{D=JC[,37]},  \textit{X=JC[,2:29]} (as columns 2 to 29 contain the pretreatment covariates), and  \textit{Y=JC[,46]}. Next, we feed these variables into the \textit{identificationDML} command, a DML-based test that by default applies lasso regression to control for $X$ and $D$ in a data-driven way. We store the results in an \textsf{R} object named \textit{output} and inspect the p-value of the test by calling \textit{output\$pval}. The box below provides the \textsf{R} code for each step.
\begin{tcolorbox}
	\begin{lstlisting}
library(causalweight)                         # load causalweight package 
data(JC)                                      # load JC data
Z=JC[,1]                                      # define instrument (assignment)
D=JC[,37]                                     # define treatment (training) 
X=JC[,2:29]                                   # define covariates 
Y=JC[,46]                                     # define outcome (health state) 
output=identificationDML(y=Y, d=D, x=X, z=Z)  # run identification test
output$pval                                   # p-value of the test
	\end{lstlisting}
\end{tcolorbox}
The test yields a p-value of  $0.26$ (or 26\%), which indicates that the null hypothesis of conditional mean independence between $Z$ and $Y$ given $D$ and $X$ cannot be rejected at conventional levels of statistical significance. For this reason, we do not find compelling statistical evidence for a violation of the selection-on-observables assumption concerning $D$ or the instrument validity of $Z$ when controlling for $X$. This suggests that the baseline covariates might be sufficiently rich to control for confounders jointly affecting $D$ and $Y$.

\begin{figure}[!htp]
	\centering \caption{\label{InvariantCausal}  Invariant causal prediction}
	\begin{center}
		\centering 	\includegraphics[scale=0.5]{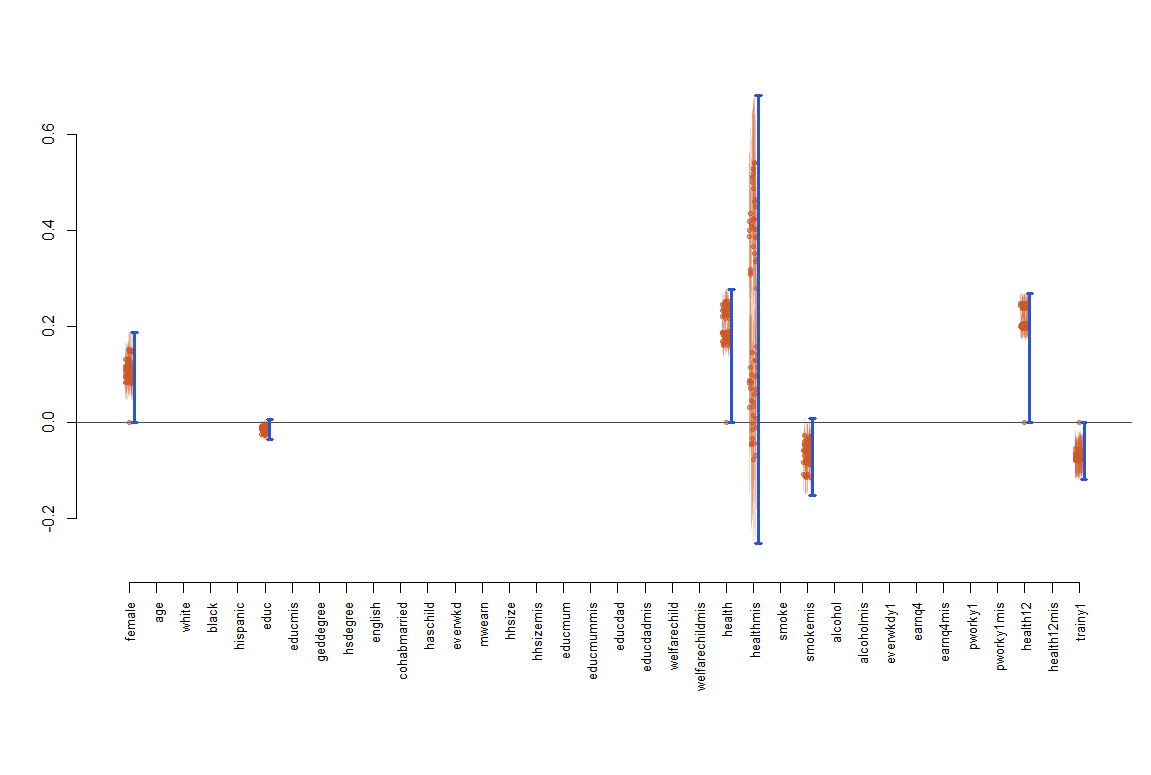}
	\end{center}
\end{figure}

In a next step, we load the \textit{InvariantCausalPrediction} package developed by \cite*{Meinshausen2019Invariant} and apply the algorithm of \cite*{peters2015causal} to identify sets of observed variables that entail invariant estimations of the conditional mean outcome when assuming a linear outcome model. We now consider all variables in the \textit{JC} data that were measured either before or during the first year after the random assignment to Job Corps as candidate treatment variables, namely, columns 2 to 37 of the data set. To this end, we define \textit{X=as.matrix(JC[,c(2:37)])}, where the \textit{as.matrix} command converts our data frame into a numeric matrix, as required by the algorithm. We note that \textit{X} includes pretreatment covariates as well as training participation and covariates measured in the first year after program assignment. As before, random assignment to Job Corps  is used as the instrument (\textit{Z}), which presumably does not directly affect the health outcome \textit{Y}, but only indirectly through elements in \textit{X} (in particular, training participation). We feed \textit{X}, \textit{Y}, \textit{Z} into the \textit{ICP} command to execute the algorithm and set the argument \textit{alpha=0.05} for testing the impact of candidate treatments at the 5\% level of significance (including a correction for multiple hypothesis testing). We save the results in an object named \textit{output} and wrap the latter by the \textit{plot} command to plot the results.  See the box below for the \textsf{R} code of the various steps. 
\begin{tcolorbox}
	\begin{lstlisting}
library(InvariantCausalPrediction)      # load InvariantCausalPrediction package
X=as.matrix(JC[,c(2:37)])               # observables (candidate treatments)
output=ICP(X=X,Y=Y,ExpInd=Z,alpha=0.05) # algorithm invariant causal prediction
plot(output)                            # plot results
	\end{lstlisting}
\end{tcolorbox}
Running the code generates the graph in figure \ref{InvariantCausal}, which conveys information regarding the selection of variables in \textit{X} as regressors in various outcome regression models. Each of these regression models adheres to the criterion that the estimation of the conditional mean outcome ($E[Y|\tilde{D}]$) based on the respective set of selected regressors ($\tilde{D}$) is rather invariant across values of the instrument \textit{Z}. In this graph, the dots correspond to the estimates of the regression coefficients (y-axis) for different variables coming from \textit{X} (x-axis) for each of the models that meet the invariance criterion. The slim vertical bars are the 95\% confidence intervals of the estimated coefficients. In contrast, the fatter vertical bars correspond to the union (or combination) of the confidence intervals across all models satisfying the invariance condition, i.e., they are derived from the maximum upper and minimum lower bounds of any confidence interval across all relevant models. 

Variables without any coefficient estimates and confidence intervals are never selected to be in the set $\tilde{D}$. Among the variables having nonzero coefficients (and thus, nonzero effects on the health outcome) in models satisfying the invariance criterion are gender, education, health at baseline and after the first year, missing dummies for smoking behavior and health at baseline, and training participation in the first year. However, for any of those variables, the union of the confidence intervals includes a zero effect. Nevertheless, we can observe, for instance, that the coefficients on training participation (\textit{trainy1}) are often negative and never positive, suggesting a health-increasing effect (due to the inverse coding of the outcome). As already mentioned, more instruments (particularly those with large first-stage effects on elements in \textit{X}) may be helpful in reducing estimation uncertainty and thus the length of the confidence intervals. Put differently, having more and stronger instruments reduces the ambiguity about which and how candidate treatments affect the outcome of interest. 

\section{Conclusion}\label{conclusion}

The application of causal discovery, aimed at learning causal relationships among multiple variables in a data-driven manner, remains so far relatively uncommon in economics and social sciences. One likely reason is that assumptions and causal setups that permit unambiguous learning of causality in entire systems of variables appear typically too restrictive to match real-world complexity. Another reason is that rather than understanding causality among many variables, researchers frequently focus on studying the effects of specific treatment variables based on more narrowly defined (but depending on the empirical context still challenging) assumptions. Even if learning causal structures in complex systems might be a too ambitious goal, the concepts of causal discovery can nevertheless be useful for somewhat more modest, but from an economic or scoial science perspective yet interesting causal problems. For instance, they can help determine the set of treatments directly impacting a specific outcome or for deriving and verifying conditions that (when tested and not rejected in large samples) imply the identification of treatment effects in observational data. For this reason, this study provided an introduction to fundamental concepts of causal discovery such as d-separation, causal faithfulness, Markov equivalence, or the back-door and front-door criteria, along with several empirical illustrations using the statistical software (\textsf{R}).

\pagebreak

\setlength\baselineskip{14.0pt}
\bibliographystyle{chicago}
{\footnotesize
	\bibliography{research_second}
}

\pagebreak

\end{document}